\documentclass[preprint]{revtex4}
\begin{document}
\title{Three classes of nonextensive entropies\\
characterized by Shannon additivity and pseudoadditivity}
\author{Hiroki Suyari}
\affiliation{%
Department of Information and Image Sciences,
Faculty of Engineering, Chiba University\\
1-33, Yayoi-cho, Inage-ku, Chiba-shi, Chiba, 263-8522 Japan
}%
\email{suyari@tj.chiba-u.ac.jp}
\date{\today\\ \bigskip\bigskip\bigskip\bigskip\bigskip}

\begin{abstract}
\bigskip
Nonextensive entropies are divided into three classes,
each of which is characterized
by Shannon additivity and pseudoadditivity.
One of the three classes has properties of both additivities.
The remaining classes have only one property of the two additivities,
respectively.
An example of nonextensive entropy is shown concretely for each class.
In particular,
one class is found to consist of only the Tsallis entropy.
More precisely,
the Tsallis entropy is proved to be uniquely
determined by only these two additivities.
The present classification using these two distinct additivities
reveals unique characteristics of the Tsallis entropy.
\end{abstract}
\maketitle

\section{Introduction}
Shannon additivity and pseudoadditivity
are characteristic properties of Tsallis entropy \cite{AO01}.
Both of these additivities are nonextensive generalizations of the standard
additivity in extensive systems.
Although these properties appear to be similar
in the sense that both additivities are given by the formulations of the entropy
for the composite systems, $A$ and $B$,
they are actually different.
When two systems, $A$ and $B$, are mutually independent,
in extensive systems, the Shannon additivity coincides with the pseudoadditivity, but
in nonextensive systems, this coincidence generally does not hold.

In the present paper,
these two additivities are applied to characterization of three classes
of nonextensive entropies.
One of the three classes is characterized by satisfying both of these additivities,
and the remaining two classes are characterized by satisfying only one 
or the other of the two additivities.
An example of nonextensive entropy is shown concretely for each class.
In addition,
the Tsallis entropy is uniquely determined by only these two additivities.
In the present paper, both the original Tsallis entropy
and the normalized Tsallis entropy are discussed.

\section{Shannon additivity and pseudoadditivity of the original Tsallis Entropy}

The {\it original Tsallis entropy} $S_q$ \cite{Ts88} is given by
\begin{equation}
S_q\left( {p_1,\cdots ,p_n} \right):={{1-\sum\limits_{i=1}^n {p_i^q}} \over {q-1}}
\label{Tsallisentropy}
\end{equation}
where $q\in R^+$.
This nonextensive entropy is one-parameter generalization of the Shannon entropy
in the sense that
\begin{equation}
\mathop {\lim }\limits_{q\to 1}S_q=S_1:=-\sum\limits_{i=1}^n {p_i\ln p_i}.
\label{Shannonentropy}
\end{equation}
The Shannon additivity \cite{CT91} and the pseudoadditivity \cite{AO01} of the original
Tsallis entropy are given by followings:
\begin{description}
\item{(1) Shannon additivity:}
\begin{eqnarray}
&&\forall i=1,\cdots,n,\forall j=1,\cdots,m_i:
p_{ij}\ge 0,\quad p_i=\sum\limits_{j=1}^{m_i} {p_{ij}},\quad
\sum\limits_{i=1}^n {p_i}=1,\\
&&S_q\left( {p_{11},\cdots,p_{nm_n}} \right)
=S_q\left( {p_1,\cdots,p_n}\right)
+\sum\limits_{i=1}^n {p^q_iS_q\!\left( {{{p_{i1}}
\over {p_i}},\cdots,{{p_{im_i}}\over {p_i}}} \right)}.
\label{Shannonadditivity}
\end{eqnarray}
\item{(2) pseudoadditivity:}
\begin{equation}
S_q\left( A,B \right)=S_q\left( A \right)+S_q\left( B \right)+
\left( 1-q \right)S_q\left( A \right)S_q\left( B \right)
\label{pseudoadditivity}
\end{equation}
where $A$ and $B$ are mutually independent finite event systems:
\begin{equation}
A=\left( {\matrix{{A_1}&\cdots &{A_n}\cr
{p_1^A}&\cdots &{p_n^A}\cr
}} \right),\;\quad B=\left( {\matrix{{B_1}&\cdots &{B_m}\cr
{p_1^B}&\cdots &{p_m^B}\cr
}} \right).
\label{ABsystem}
\end{equation}
\end{description}

These two additivities, Eqs.(\ref{Shannonadditivity}) and (\ref{pseudoadditivity}),
can be proven in a straightforward manner using the definition 
given in Eq.(\ref{Tsallisentropy}).
If the condition of independence given by Eq.(\ref{ABsystem}):
\begin{equation}
\forall i=1,\cdots ,n,\;\forall j=1,\cdots ,m\;\;:\;\;
p_{ij}=p_i\cdot p_j,
\end{equation}
is applied to the Shannon additivity given by Eq.(\ref{Shannonadditivity}),
then the equation is simplified to
\begin{equation}
\hspace{-0.5cm}S_q\!\left( {p_{11},\;\cdots ,\;p_{nm}} \right)
=S_q\!\left( {p_1,\;\cdots ,\;p_n}\right)
+\left(\sum\limits_{i=1}^n p_i^q\right)
S_q\!\left( {p_1,\;\cdots,\;p_m}\right).
\label{reducedShannonadditivity}
\end{equation}
Here, we take $m:=m_i$ for all $i=1,\cdots ,n$.
In terms of the following notations:
\begin{eqnarray}
&&\forall i=1,\cdots ,n:p_i:=p_i^A,\;\;
\forall j=1,\cdots ,m:p_j:=p_j^B,\;\\
&&S_q\left( A \right)=S_q\left( {p_1,\;\cdots ,\;p_n}\right),\;\;
S_q\left( B \right)=S_q\left( {p_1,\;\cdots ,\;p_m}\right)
\end{eqnarray}
the obtained Shannon additivity, Eq.(\ref{reducedShannonadditivity}), is written as
\begin{equation}
S_q\left( {A,B} \right)
=S_q\left( A \right)+\left( {\sum\limits_{i=1}^n p_i^q}\right)
S_q\left( B \right).
\label{simpleadditivity}
\end{equation}
Note that in the case of $q=1$ (extensive systems),
the Shannon additivity finally reaches
\begin{equation}
S_1\left( {A,B} \right)=S_1\left( A \right)+S_1\left( B \right),
\label{standardadditivity}
\end{equation}
which coincides with the pseudoadditivity of Eq.(\ref{pseudoadditivity}).
In this way, 
when two systems, $A$ and $B$, are mutually independent in the extensive systems,
the Shannon additivity coincides with the pseudoadditivity.

However, under the above condition of independence, Eq.(\ref{ABsystem}),
in nonextensive systems, 
these two additivities differ from each other
in the sense that one additivity cannot be derived from the other.
Comparing Eqs.(\ref{pseudoadditivity}) and (\ref{simpleadditivity}),
these additivities do not coincide with each other.

Therefore, the nonextensive entropies are divided into 
the following three classes:
\begin{description}
\item{Class 1:} Nonextensive entropy $S_q^{(1)}$ satisfying both additivities

\item{Class 2:} Nonextensive entropy $S_q^{(2)}$ satisfying only the Shannon
additivity

\item{Class 3:} Nonextensive entropy $S_q^{(3)}$ satisfying only the
pseudoadditivity

\end{description}
Here, we need a condition that all nonextensive entropies should satisfy 
Eq.(\ref{Shannonentropy}).

Examples of each class of nonextensive entropy will be shown concretely
in order to reveal the difference between the two additivities.

First, we uniquely derive the nonextensive entropy $S_q^{(1)}$ of Class 1
from only these two additivities,
and the obtained entropy is found to coincide with the original Tsallis entropy.

If two systems, $A$ and $B$, are mutually independent in the nonextensive systems,
then the two additivities are given by Eqs.(\ref{pseudoadditivity})
and (\ref{simpleadditivity}), respectively.
Eliminating $S_q\left( {A,B} \right)$ from both Eqs.(\ref{pseudoadditivity})
and (\ref{simpleadditivity}) yields
\begin{equation}
S_q^{(1)}\!\left( A \right)+S_q^{(1)}\!\left( B \right)+
\left( 1-q \right)S_q^{(1)}\!\left( A \right)S_q\left( B \right)
=S_q^{(1)}\!\left( A \right)+\left( {\sum\limits_{i=1}^n p_i^q}\right)
S_q^{(1)}\!\left( B \right).
\label{eliminate0}
\end{equation}
Therefore, the original Tsallis entropy, Eq.(\ref{Tsallisentropy}),
is directly derived from Eq.(\ref{eliminate0}), as follows:
\begin{equation}
S_q^{(1)}\!\left( A \right)
={{1-\sum\limits_{i=1}^n {p_i^q}} \over {q-1}}
\end{equation}

Note that in this derivation only the Shannon additivity and pseudoadditivity
are used.
In other words, these two additivities uniquely determine
the original Tsallis entropy.

We must note here that our derivation is not consistent 
with the result reported in \cite{Sa97}.
As reported in \cite{Sa97}, the four conditions: (i) continuity, (ii) increasing
monotonicity, (iii) pseudoadditivity, Eq.(\ref{pseudoadditivity}), and
(iv) Shannon additivity, Eq.(\ref{Shannonadditivity}), are given
as axioms of the original Tsallis entropy and the uniqueness theorem
thereof is proven.
Note that the Shannon additivity in \cite{Sa97} is a special case of
Eq.(\ref{Shannonadditivity}).
The generalization of the Shannon additivity in \cite{Sa97}
to Eq.(\ref{Shannonadditivity}) is straightforward \cite{Ts95,AO01}.
In the present derivation, we uniquely determine the Tsallis entropy
using only two axioms, (iii) and (iv) described above.
Therefore, the above four conditions in \cite{Sa97} are {\it redundant}
axioms of the Tsallis entropy.
Thus, these two additivities become self-consistent axioms of the original Tsallis
entropy, but complete parallelism does not exist
between the Shannon-Khinchin axioms and the two additivities.

Next, we consider the following example of the nonextensive entropy $S_q^{(2)}$
of Class 2,
\begin{equation}
S_q^{(2)}\!\left( {p_1,\cdots ,p_n} \right)
:={{1-\sum\limits_{i=1}^n {p_i^q}} \over {\varphi\left( q \right)}}
\label{class2entropy}
\end{equation}
where $\varphi \left( q \right)$ is a differentiable function
with respect to any $q\in R^+$, satisfying
\begin{equation}
\mathop {\lim }\limits_{q\to 1}
{{d\varphi \left( q \right)} \over {dq}}=1,\,\quad
\mathop {\lim }\limits_{q\to 1}
\varphi \left( q \right)=\varphi \left( 1\right)=0,\,\quad
\varphi \left( q \right)\ne0\; \left( {q\!\ne\!1} \right),\,\quad
\hbox{and}\quad\varphi \left( q \right)\ne q-1.
\label{conditionvarphi2}
\end{equation}
The difference in $S_q^{(2)}$ from that of 
the original Tsallis entropy, Eq.(\ref{Tsallisentropy}),
is only the denominator ${\varphi\left( q \right)}$.
Hence, the last condition of Eq.(\ref{conditionvarphi2}) is required
in order to belong to Class 2.
The following equation for $\varphi \left( q \right)$ is an example 
satisfying all conditions of Eq.(\ref{conditionvarphi2}), 
\begin{equation}
\varphi \left( q \right)={{\left( {q-1} \right)\left( {q^2+1} \right)}
\over 2}.
\label{examplevarphi}
\end{equation}
We must confirm that the nonextensive entropy $S_q^{(2)}$
belongs to Class 2.
Using l'Hospital's rule, we have
\begin{equation}
\mathop {\lim }\limits_{q\to 1}S_q^{(2)}=\mathop {\lim }\limits_{q\to
1}{{1-\sum\limits_{i=1}^n {p_i^q}} \over {\varphi \left( q \right)}}
={\mathop {\lim}\limits_{q\to 1}}
{{-\sum\limits_{i=1}^n {p_i^q\ln p_i}} \over
{{d\varphi\left( q\right)} \over {dq}}}
=-\sum\limits_{i=1}^n {p_i\ln p_i}=S_1.
\end{equation}
Thus, $S_q^{(2)}$ satisfies 
condition of Eq.(\ref{Shannonentropy}).
$S_q^{(2)}$ is found to have the property of
the Shannon additivity, Eq.(\ref{Shannonadditivity}), as follows:
\begin{eqnarray}
S_q^{(2)}\!\left( {p_1,\cdots,p_n}\right)
+\sum\limits_{i=1}^n {p^q_iS_q^{(2)}\!\left( {{{p_{i1}}
\over {p_i}},\cdots,{{p_{im_i}}\over {p_i}}} \right)}
&=&{{1-\sum\limits_{i=1}^n {p_i^q}} \over {\varphi\left( q \right)}}
+\sum\limits_{i=1}^n {\left( {p_i^q\cdot {{1-\sum\limits_{j=1}^{m_i} {\left( {{{p_{ij}}
\over {p_i}}} \right)^q}} \over {\varphi \left( q \right)}}} \right)}\nonumber\\
&=&S_q^{(2)}\!\left( {p_{11},\cdots,p_{nm_n}} \right).\nonumber
\end{eqnarray}
However, $S_q^{(2)}$ does {\it not} satisfy the
pseudoadditivity of Eq.(\ref{pseudoadditivity}).
\begin{equation}
S_q^{(2)}\!\left( A,B \right)\ne S_q^{(2)}\!\left( A \right)+S_q^{(2)}\!\left( B \right)+
\left( 1-q \right)S_q^{(2)}\!\left( A \right)S_q^{(2)}\!\left( B \right)
\end{equation}
Thus, $S_q^{(2)}$ belongs to Class 2.

Finally, we consider the nonextensive entropies of Class 3 in the following
example:
%
\begin{equation}
S_q^{\left( 3 \right)}\!\left( {p_1,\cdots,p_n}\right)
:={{\sum\limits_{i=1}^n{p_i^{q^{-1}}\left( {p_i^{q-1}-1}
\right)}} \over {\left( {1-q} \right)\sum\limits_{i=1}^n
{p_i^{q^{-1}}}}}={{\sum\limits_{i=1}^n {p_i^{q+q^{-1}-1}}-\sum\limits_{i=1}^n
{p_i^{q^{-1}}}} \over {\left( {1-q} \right)\sum\limits_{i=1}^n {p_i^{q^{-1}}}}}
\label{class3entropy}
\end{equation}
%
Again, using l'Hospital's rule, we have
\begin{eqnarray}
\mathop {\lim }\limits_{q\to 1}S_q^{\left( 3 \right)}
&=&\mathop {\lim }\limits_{q\to
1}{{\sum\limits_{i=1}^n {p_i^{q+q^{-1}-1}}-\sum\limits_{i=1}^n {p_i^{q^{-1}}}} \over
{\left( {1-q} \right)\sum\limits_{i=1}^n {p_i^{q^{-1}}}}}
={\mathop {\lim }\limits_{q\to 1}}
{{ {\left( {1-{1 \over {q^2}}}
\right)\sum\limits_{i=1}^n {p_i^q\ln p_i}+{1 \over {q^2}}\sum\limits_{i=1}^n
{p_i^{q^{-1}}\ln p_i}} } \over {
{-\sum\limits_{i=1}^n {p_i^{q^{-1}}}-\left( {1-q} \right){1
\over {q^2}}\sum\limits_{i=1}^n {p_i^{q^{-1}}\ln p_i}} }}\nonumber\\
&=&{{\sum\limits_{i=1}^n {p_i\ln p_i}} \over {-1}}=-\sum\limits_{i=1}^n {p_i\ln
p_i}=S_1.
\end{eqnarray}
Thus, $S_q^{(3)}$ satisfies 
condition of Eq.(\ref{Shannonentropy}).
$S_q^{(3)}$ can be easily found to satisfy
the pseudoadditivity of Eq.(\ref{pseudoadditivity}).
\begin{equation}
S_q^{(3)}\!\left( A,B \right)
=S_q^{(3)}\!\left( A \right)+S_q^{(3)}\!\left( B \right)+
\left( 1-q \right)S_q^{(3)}\!\left( A \right)S_q^{(3)}\!\left( B \right)
\end{equation}
However, the Shannon additivity of Eq.(\ref{Shannonadditivity})
does not hold for $S_q^{(3)}$.
%
%
Accordingly, $S_q^{(3)}$ belongs to Class 3.
\section{Shannon additivity and pseudoadditivity of the normalized Tsallis Entropy}

The {\it normalized Tsallis entropy} $\hat S_q$ was first introduced
in \cite{LV98} as one candidate of the generalized
nonextensive entropies.
The requirement for the normalized Tsallis entropy is presented
based on the principle of the form invariance of Kullback-Leibler entropy \cite{RA99}.
or on that of the pseudoadditivity \cite{Su01a}, respectively.
The {\it normalized Tsallis entropy} $\hat S_q$ \cite{LV98,RA99,Su01a} is given by
\begin{equation}
\hat S_q\left( {p_1,\cdots ,p_n} \right):=
{{1-\sum\limits_{i=1}^n {p_i^q}} \over {\left( {q-1} \right)\sum\limits_{j=1}^n
{p_j^q}}}
\label{n_Tsallisentropy}
\end{equation}
where $q\in R^+$.
This nonextensive entropy is also a one-parameter generalization of the Shannon entropy
in the sense that
\begin{equation}
\mathop {\lim }\limits_{q\to 1}{\hat S_q}=S_1=-\sum\limits_{i=1}^n {p_i\ln p_i}.
\label{n_Shannonentropy}
\end{equation}
The Shannon additivity and the pseudoadditivity \cite{AO01} of the normalized
Tsallis entropy are given by the followings:
\begin{description}
\item{(1) Shannon additivity:}
\begin{eqnarray}
&&\forall i=1,\cdots,n,\forall j=1,\cdots,m_i:
p_{ij}\ge 0,\quad p_i=\sum\limits_{j=1}^{m_i} {p_{ij}},\quad
\sum\limits_{i=1}^n {p_i}=1,\\
\lefteqn{\left( {\sum\limits_{i=1}^n {\sum\limits_{j=1}^{m_i} {p_{ij}^q}}} \right)
{\hat S}_q\!\left({p_{11},\;\cdots ,\;p_{1m_1},\;\cdots ,\;p_{n1},\;\cdots
,\;p_{nm_n}}
\right)}
\quad\quad\nonumber\\
&&=\left( {\sum\limits_{i=1}^n {p_i^q}} \right){\hat S}_q\!\left( {p_1,\;\cdots ,\;p_n}
\right)+\sum\limits_{i=1}^n {\sum\limits_{j=1}^{m_i} {p_{ij}^q{\hat S}_q\!
\left({{{p_{i1}} \over {p_i}},\;\cdots ,\;{{p_{im_i}} \over {p_i}}} \right)}}
\label{n_Shannonadditivity}
\end{eqnarray}
\item{(2) pseudoadditivity:}
\begin{equation}
{\hat S_q}\left( A,B \right)={\hat S_q}\left( A \right)+{\hat S_q}\left( B \right)+
\left( q-1 \right){\hat S_q}\left( A \right){\hat S_q}\left( B \right)
\label{n_pseudoadditivity}
\end{equation}
where $A$ and $B$ are mutually independent finite event systems
given by Eq.(\ref{ABsystem}).
\end{description}

The two additivities given by Eqs.(\ref{n_Shannonadditivity}) and
(\ref{n_pseudoadditivity}) are proven using the definition given in
Eq.(\ref{n_Tsallisentropy}).
Note that the difference between the two Shannon additivities
given by Eqs.(\ref{Shannonadditivity}) and (\ref{n_Shannonadditivity})
is due to the normalization of the Tsallis entropy.
The two Tsallis entropies $S_q$ and $\hat S_q$ defined by
Eqs.(\ref{Tsallisentropy}) and (\ref{n_Tsallisentropy}) yield the following
relation:
\begin{equation}
S_q\!\left( {p_1,\;\cdots ,\;p_n}\right)
=\left({\sum\limits_{j=1}^n{p_j^q}}\right)
{\hat S_q}\!\left( {p_1,\;\cdots ,\;p_n}\right).
\label{n_relation}
\end{equation}
Thus, substituting Eq.(\ref{n_relation}) into Eq.(\ref{Shannonadditivity})
yields the Shannon additivity of Eq.(\ref{n_Shannonadditivity})
for the normalized Tsallis entropy $\hat S_q$.
Another difference in pseudoadditivity is that
the coefficient \lq\lq $1-q$" in front of the cross term in the
right-hand side of the pseudoadditivity of Eq.(\ref{pseudoadditivity})
is the inverse of \lq\lq $q-1$" in Eq.(\ref{n_pseudoadditivity}).
The reason for this difference is
the normalization of the original Tsallis entropy,
which is discussed in detail in our paper \cite{Su01a}.

Similar to the case of the original Tsallis entropy,
the condition of independence given by Eq.(\ref{ABsystem}) is applied to
the Shannon additivity of Eq.(\ref{n_Shannonadditivity}),
which yields
\begin{equation}
\left( {\sum\limits_{j=1}^m {p_j^q}} \right)\hat S_q\!\left( {A,B} \right)
=\hat S_q\!\left(A \right)
+\left( {\sum\limits_{j=1}^m {p_j^q}} \right)\hat S_q\!\left(B \right).
\label{n_simpleadditivity}
\end{equation}
Here, we take $m:=m_i$ for all $i=1,\cdots ,n$.
Note that in the case of $q=1$ (extensive systems) 
the Shannon additivity becomes the standard additivity 
of Eq.(\ref{standardadditivity}),
which coincides with the pseudoadditivity of 
Eq.(\ref{n_pseudoadditivity}).
In this way, 
when two systems, $A$ and $B$, are mutually independent in the extensive systems,
the Shannon additivity coincides with the pseudoadditivity.

However, under the above condition of independence
in the nonextensive systems, 
these two additivities differ from each other
in the sense that one additivity cannot be derived from the other.
Comparing Eqs.(\ref{n_pseudoadditivity}) and (\ref{n_simpleadditivity}),
these additivities do not coincide with each other.

Therefore, the normalized nonextensive entropies are divided into 
the following three classes:
\begin{description}
\item{Class 1:} Nonextensive entropies $\hat S_q^{(1)}$
satisfying both additivities

\item{Class 2:} Nonextensive entropies $\hat S_q^{(2)}$
satisfying only the Shannon additivity

\item{Class 3:} Nonextensive entropies $\hat S_q^{(3)}$
satisfying only the pseudoadditivity

\end{description}
Here, we need a condition that all nonextensive entropies should satisfy 
Eq.(\ref{n_Shannonentropy}).

Examples of each class of nonextensive entropy in each class will be
shown concretely in order to reveal the difference between the two additivities.

First, we uniquely derive the nonextensive entropy $\hat S_q^{(1)}$ of Class 1
as in the previous section.
If two systems, $A$ and $B$, are mutually independent in the extensive systems, then
the two additivities are given by Eqs.(\ref{n_pseudoadditivity}) and
(\ref{n_simpleadditivity}), respectively.
Eliminating $S_q\left( {A,B} \right)$ from
both Eqs.(\ref{n_pseudoadditivity}) and (\ref{n_simpleadditivity}) yields
\begin{equation}
\left( {\sum\limits_{j=1}^m {p_j^q}} \right)\left\{ {\hat S_q\!\left( A \right)+\hat
S_q\!\left( B \right)+\left( {q-1} \right)\hat S_q\!\left( A \right)\hat S_q\!\left( B
\right)} \right\}=\hat S_q\!\left( A \right)+\left( {\sum\limits_{j=1}^m {p_j^q}}
\right)\hat S_q\!\left( B \right).
\label{eliminate1}
\end{equation}
Therefore, the normalized Tsallis entropy of Eq.(\ref{n_Tsallisentropy})
is directly derived from Eq.(\ref{eliminate1}) as follows:
\begin{equation}
\hat S_q^{(1)}\!\left( B \right)
={{1-\sum\limits_{i=1}^m {p_i^q}} \over {\left( {q-1} \right)\sum\limits_{j=1}^m
{p_j^q}}}
\end{equation}

Note that in this derivation
only the Shannon additivity and pseudoadditivity are used.
In other words,
these two additivities uniquely determine the normalized Tsallis entropy,
similar to the case of the original Tsallis entropy.

Next, we consider the following example of the nonextensive entropy $\hat S_q^{(2)}$
of Class 2,
\begin{equation}
\hat S_q^{(2)}\!\left( {p_1,\cdots ,p_n} \right)
:={{1-\sum\limits_{i=1}^n {p_i^q}} \over {\varphi\left( q \right)
{\sum\limits_{j=1}^n{p_j^q}}}}
\label{n_class2entropy}
\end{equation}
where $\varphi \left( q \right)$ is a differentiable function
with respect to any $q\in R^+$, satisfying the conditions 
of Eq.(\ref{conditionvarphi2}).
Here, we must verify that the nonextensive entropy $\hat S_q^{(2)}$
belongs to Class 2.
Using l'Hospital's rule, we have
\begin{equation}
\mathop {\lim }\limits_{q\to 1}\hat S_q^{(2)}=\mathop {\lim
}\limits_{q\to 1}
{{1-\sum\limits_{i=1}^n {p_i^q}} \over {\varphi\left( q \right)
{\sum\limits_{j=1}^n{p_j^q}}}}
={\mathop {\lim }\limits_{q\to 1}}
{{{-\sum\limits_{i=1}^n {p_i^q\ln p_i}} } \over 
{ {{{d\varphi \left( q \right)} \over
{dq}}\sum\limits_{j=1}^n {p_j^q}+\varphi
\left( q \right)\sum\limits_{j=1}^n {p_j^q\ln p_j}} }}
=-\sum\limits_{i=1}^n {p_i\ln p_i}=S_1.
\end{equation}
Thus, $\hat S_q^{(2)}$ satisfies 
the condition of Eq.(\ref{n_Shannonentropy}).
$\hat S_q^{(2)}$ is found to have the property of 
the Shannon additivity Eq.(\ref{n_Shannonadditivity}).

However, $\hat S_q^{(2)}$ does {\it not} satisfy the
pseudoadditivity of Eq.(\ref{n_pseudoadditivity})
\begin{equation}
{\hat S}_q^{(2)}\!\left( A,B \right)\ne {\hat S}_q^{(2)}\!\left( A \right)
+{\hat S}_q^{(2)}\!\left( B\right)+
\left( q-1 \right){\hat S}_q^{(2)}\!\left( A \right){\hat S}_q^{(2)}\!\left( B \right).
\end{equation}
Thus, ${\hat S}_q^{(2)}$ belongs to Class 2.

Finally, we consider the nonextensive entropies of Class 3
in the following example:
%
\begin{equation}
\hat S_q^{\left( 3 \right)}
:={{\sum\limits_{i=1}^n {p_i^{{{q^2+1} \over 2}}
\left({p_i^{1-q}-1} \right)}} \over {\left( {q-1} \right)\sum\limits_{i=1}^n
{p_i^{{{q^2+1}\over 2}}}}}
={{\sum\limits_{i=1}^n {p_i^{{{q^2-2q+3} \over
2}}}-\sum\limits_{i=1}^n {p_i^{{{q^2+1} \over 2}}}} \over {\left( {q-1}
\right)\sum\limits_{i=1}^n {p_i^{{{q^2+1}
\over 2}}}}}
\label{n_class3entropy}
\end{equation}
%
Again, using l'Hospital rule, we have
\begin{eqnarray}
\mathop {\lim }\limits_{q\to 1}\hat S_q^{\left( 3 \right)}
&=&\mathop {\lim }\limits_{q\to
1}{{\sum\limits_{i=1}^n {p_i^{{{q^2-2q+3} \over 2}}}-\sum\limits_{i=1}^n {p_i^{{{q^2+1}
\over 2}}}} \over {\left( {q-1} \right)\sum\limits_{j=1}^n {p_j^{{{q^2+1} \over
2}}}}}
={\mathop {\lim }\limits_{q\to 1}}
{{ {\left( {q-1} \right)\sum\limits_{i=1}^n
{p_i^{{{q^2-2q+3} \over 2}}\ln p_i}-q\sum\limits_{i=1}^n {p_i^{{{q^2+1} \over 2}}\ln
p_i}} } \over { {\sum\limits_{j=1}^n
{p_j^{{{q^2+1} \over 2}}}+q\left( {q-1} \right)\sum\limits_{j=1}^n {p_j^{{{q^2+1} \over
2}}\ln p_j}} }}\nonumber\\
&=&{{-\sum\limits_{i=1}^n {p_i\ln p_i}} \over 1}=-\sum\limits_{i=1}^n {p_i\ln p_i}=S_1
\end{eqnarray}
Thus, $\hat S_q^{(3)}$ satisfies 
the condition of Eq.(\ref{n_Shannonentropy}).
In addition, ${\hat S}_q^{(3)}$ is verified to have the
property of pseudoadditivity of Eq.(\ref{n_pseudoadditivity}).
\begin{equation}
\hat S_q^{\left( 3 \right)}\!\left( {A,B} \right)
=\hat S_q^{\left( 3 \right)}\!\left( A \right)
+\hat S_q^{\left( 3 \right)}\!\left( B\right)
+\left( {q-1} \right)\hat S_q^{\left( 3\right)}\!\left( A \right)
\hat S_q^{\left( 3\right)}\!\left( B \right)
\end{equation}
However, the Shannon additivity of Eq.(\ref{n_Shannonadditivity})
does not hold for $\hat S_q^{(3)}$.
%
%
Accordingly, $\hat S_q^{(3)}$ belongs to Class 3.

\section{Conclusion}

We have shown that
the nonextensive entropies can be divided into three classes
which are characterized by the Shannon additivity and the pseudoadditivity.
An example of each class is shown concretely and
these two distinct additivities reveal the following peculiarities
of the Tsallis entropy.

First, the Tsallis entropy is uniquely determined 
by only these two additivities.
Thus, these two additivities constitute the axioms of the Tsallis entropy.
However, these axioms are {\it not} a naturally generalization of
the Shannon-Khinchin axioms in extensive systems
due to the lack of complete parallelism between them.

Secondly, there exist nonextensive entropies
satisfying only one of the two additivities.
Accordingly, for the construction of the general axioms for the Tsallis entropy,
unsatisfactory axioms consisting of these two additivities are observed.
In other words,
axioms that include both additivities are redundant.


\end{document}